\documentclass[a4paper]{article}

\usepackage{INTERSPEECH2022}

\usepackage{graphicx}
\usepackage{amssymb,amsmath,bm}
\usepackage{textcomp}
\usepackage{url}
\usepackage[utf8]{inputenc}
\usepackage{CJKutf8}
\usepackage{multirow}
\usepackage{comment}

\newcommand{\jp}[1]{ \begin{CJK}{UTF8}{ipxm}#1\end{CJK} }

\ninept

\title{STUDIES: Corpus of Japanese Empathetic Dialogue Speech \\ Towards Friendly Voice Agent}

\makeatletter
\def\name#1{\gdef\@name{#1\\}}
\makeatother
\name{{\em Yuki Saito$^{1}$, Yuto Nishimura$^{1}$, Shinnosuke Takamichi$^1$, Kentaro Tachibana$^2$, and Hiroshi Saruwatari$^1$}}

\address{
$^1$The University of Tokyo, Japan, $^2$LINE Corp., Japan.
}
\email{
yuuki\_saito@ipc.i.u-tokyo.ac.jp
}

\begin{document}
\setlength{\abovedisplayskip}{5pt} 
\setlength{\belowdisplayskip}{5pt} 
\setlength\floatsep{8pt} 
\setlength\intextsep{8pt} 
\setlength\textfloatsep{8pt} 
\setlength{\dbltextfloatsep}{8pt} 
\setlength{\dblfloatsep}{5pt}

\maketitle

\begin{abstract}
We present \textit{STUDIES}, a new speech corpus for developing a voice agent that can speak in a friendly manner. Humans naturally control their speech prosody to empathize with each other. By incorporating this ``empathetic dialogue'' behavior into a spoken dialogue system, we can develop a voice agent that can respond to a user more naturally. We designed the STUDIES corpus to include a speaker who speaks with empathy for the interlocutor's emotion explicitly. We describe our methodology to construct an empathetic dialogue speech corpus and report the analysis results of the STUDIES corpus. We conducted a text-to-speech experiment to initially investigate how we can develop more natural voice agent that can tune its speaking style corresponding to the interlocutor's emotion. The results show that the use of interlocutor's emotion label and conversational context embedding can produce speech with the same degree of naturalness as that synthesized by using the agent's emotion label. Our project page of the STUDIES corpus is http://sython.org/Corpus/STUDIES. 
\end{abstract}

\vspace{-5pt}
\section{Introduction}\label{sect:intro}
\vspace{-3pt}
Text-to-speech (TTS)~\cite{sagisaka88} is a technology for synthesizing speech from input text by using a computer, which is crucial for developing a natural voice agent that can communicate with humans. Studies on deep-learning-based TTS technologies (i.e., neural TTS) have enabled natural speech synthesis from various speakers~\cite{shen18,kim21vits}. Therefore, the trend in neural TTS research is shifting toward technologies that can synthesize human-like speech with rich expressions, such as end-to-end expressive TTS modeling~\cite{lee17nips,wang18gst,li21}. However, current TTS technologies have not yet reached the level where they can reproduce a natural communication behavior in human-to-human dialogue.

Our focus in this paper for developing more natural voice agents is {\it empathy}, which is an important factor to improve emotion perception in intimate communication, e.g., doctor--patient~\cite{hojat09} and teacher--student~\cite{warren15}. Empathy is a concept similar to but different from sympathy~\cite{dabis96}. Namely, the former is an active attempt to get inside the interlocutor, while the latter, in most cases, is passive communication where a speaker's feelings are shared or identified with the interlocutor. Empathetic language resources and language generation have been studied to implement this active behavior in computers and to develop more empathetic chatbots~\cite{rashkin19,li20emodg}. However, although speech corpora of empathetic dialogue have yet to be constructed, linguistic content and prosody of speech can be used as a means of empathy~\cite{regenbogen12}. If one develops {\it empathetic dialogue speech synthesis} technology that empathizes with the interlocutor and appropriately controls the prosody of the synthesized voice, it will be possible to develop more friendly voice agents.

\begin{table}[t]
\centering
\caption{Example of dialogue lines by the teacher and male student in our STUDIES corpus. The teacher empathizes the student who got a good score on a test.}
\label{tab:dialogue_example}
\small
\vspace{-3mm}
\begin{tabular}{p{0.125\linewidth}|p{0.15\linewidth}|p{0.55\linewidth}}
\hline
\hline
Speaker &
Emotion &
Text \\
\hline
Male student &
Happy &
\jp{先生、こんにちは！}
\newline {\footnotesize c.f.) Hello, teacher!} \\
\hline
Teacher &
Neutral &
\jp{あら、その顔。テスト良かったのかな？}
\newline {\footnotesize c.f.) Oh, did you get a good score on your test?} \\
\hline
Male student &
Happy &
\jp{当ったりー！苦手な数学と化学、点数アップしました！}
\newline {\footnotesize c.f.) Bingo! I improved my scores in math and chemistry, my two weakest subject areas!} \\
\hline
Teacher &
Happy &
\jp{おめでとう！良かったね！}
\newline {\footnotesize c.f.) Congratulations! Good for you!}\\
\hline
\hline
\end{tabular}
\end{table}

To advance empathetic dialogue speech synthesis research, we present {\it STUDIES}\footnote{It is an abbreviation of ``Stuff To Understand Dialogue Including Empathetic Speech.''}, a new open-source empathetic dialogue speech corpus for developing more friendly TTS technologies. Our STUDIES corpus contains simulated chat-dialogue uttered by three actors who act as a female teacher and her male/female students at a tutoring school. With a view to develop applications for more empathetic chat-oriented spoken dialogue systems~\cite{mctear20}, we designed this corpus to include empathetic speech acted out by the teacher attempting to motivate her students to learn and improve their enjoyment of school life, as shown in Table~\ref{tab:dialogue_example}. We describe our methodology to construct an empathetic dialogue speech corpus and present the analysis results of empathetic speech in the STUDIES corpus. We also conducted a TTS experiment using the STUDIES corpus and investigated what kind of information is effective for improving the quality of speech synthesized by empathetic dialogue speech synthesis. The contributions of this study are as follows:
    \vspace{-1mm}
    \begin{itemize} \leftskip -5mm \itemsep -0mm
        \item We construct a new corpus that includes multiple professional speakers for developing more friendly neural TTS systems. Our corpus is open-sourced for research purposes only from our project page\footnote{http://sython.org/Corpus/STUDIES}.
        \item We discuss the novel and meaningful points of the STUDIES corpus on the basis of a comparison with existing corpora.
        \item We present the results of a TTS experiment and demonstrate that interlocutor's emotion labels and conversational context embedding enable synthesizing speech with the same degree of naturalness as that synthesized by using speaker's emotion label.
    \end{itemize}
    \vspace{-2mm}

\vspace{-5pt}
\section{Corpus Construction Methodology for Empathetic Spoken Dialogue}
\vspace{-3pt}
We describe a crowdsourcing-based methodology for constructing an empathetic spoken dialogue corpus.

\vspace{-3pt}
\subsection{Dialogue scenario}\label{subsect:scenario}
\vspace{-3pt}
Common issues~\cite{zhang18acl} of chat-oriented dialogue with a specific persona include 1) inconsistent personality~\cite{li16acl} and 2) lack of long-term memory~\cite{vinyals15} (i.e., chat history).

For the first issue, we specified the persona of the female teacher (i.e., voice agent of the dialogue system) before collecting chat lines. Her persona was ``a female in her early twenties, Tokyo dialect speaker, cheerful and friendly personality, gentle tone of voice, and willing to lead the conversation to some extent,'' which was consistent throughout all dialogue lines. We did not specify any persona of the students and let the crowdworkers come up with them\footnote{We extracted the student persona information conceived by the writers of dialogue lines from the crowdsourcing outcomes and described them in ``dialog.xml,'' which is stored in each long dialogue directory.} because we assume that various users will interact with the actual voice agent. We also prepared several settings of chat situations where students can easily express their emotions (e.g., ``The teacher praises her student who recorded a good score on the test.'') and students' initial emotions. This procedure was similar to the situation-grounded dialogue collection for the empathetic text-chat corpus~\cite{rashkin19}.

For the second issue, we prepared two types of dialogues: long (10--20 turns) and short (4 turns). The former is a difficult task for both crowdworkers who make the lines and a speech synthesizer that synthesizes from the lines considering long-term history. The latter is suitable for a simplified evaluation of empathetic dialogue systems. In addition, the lines can be made easily by crowdworkers than those of long dialogues.

Our methodology is much inspired by the EmpatheticDialogues dataset~\cite{rashkin19}, i.e., a crowdsourced large-scale text-chat corpus containing conversations under a given situation description and fine-grained emotion label for each conversation. One major difference between ours and the conventional one is that we asked the crowdworkers to make {\it utterance-wise} emotion labels in addition to {\it situation-wise} ones. This is because humans can appropriately control speech prosody in accordance with the interlocutor's current emotion per utterance, even if the specific situation is given.

\vspace{-3pt}
\subsection{Crowdsourcing chat dialogue lines}\label{subsect:script_collection}
\vspace{-3pt}
We crowdsourced lines of chat dialogue between the teacher and male/female students. We adopted macrotask and microtask crowdsourcing methods~\cite{simperl15}.

\vspace{-3pt}
\subsubsection{Macrotask crowdsourcing for long-dialogue lines}
\vspace{-3pt}
Since making the long-dialogue lines requires a significant level of skill and deep understanding of scenarios, we adopted macrotask crowdsourcing, where crowdworkers are employed by clients on the basis of their skills and experiences~\cite{simperl15}. We presented the employed crowdworkers with 1) the dialogue scenario (teacher-student empathetic dialogue in a tutoring school), 2) the teacher's persona (described in Section~\ref{subsect:scenario}), 3) dialogue duration (10--20 turns), 4) interlocutor (half of the lines for the teacher and a male student, and the remainder for the teacher and a female student), and 5) the students' initial emotions. The emotions included neutral, happy, sad, and angry, which were assigned to each crowdworker in a balanced manner. Each crowdworker made a number of dialogue lines that satisfied the presented condition. We did not specify the students' persona nor which character would speak first.

\vspace{-3pt}
\subsubsection{Microtask crowdsourcing for short-dialogue lines}
\vspace{-3pt}
For collecting short-dialogue lines, we adopted microtask crowdsourcing, where any crowdworkers regardless of their skills and experiences can participate in the microtask (typically done in a few minutes) without approval by the client~\cite{simperl15}. We presented the crowdworkers with conditions similar to those in the long-dialogue case, but excluded the students' initial emotions to make the task simple. Since this type of crowdsourcing suffers from quality assurance~\cite{dawid79}, we manually revised the lines to remove sentences including typos or offensive expressions.

\vspace{-3pt}
\subsection{Voice recording}
\vspace{-3pt}
We employed three professional speakers (one man and two women) and recorded simulated chat dialogues between the teacher and male/female students in a recording studio on the basis of the collected dialogue lines. This recording process was conducted on separate days for each speaker as a countermeasure against the COVID-19 pandemic. The sampling rate was 48~kHz.

As later discussed in Section~\ref{subsect:spec}, the dialogue lines collected through our crowdsourcing were highly imbalanced regarding the number of utterances per emotion label. To deal with this issue, we additionally recorded the phoneme-balanced emotional speech of the actor who performed as the teacher for robust training of our speech synthesizer. Specifically, we asked the actor to read phonetically balanced sentences in the ITA corpus~\cite{ita} given emotion labels.

\begin{table}[t]
\centering
\caption{Number of utterances per each subset}
\vspace{-3mm}
\label{tab:utterances_per_subset}
\begin{tabular}{c|rrr|r}
\hline
\hline
Speaker & Long & Short & ITA & Total (hours) \\
\hline
Teacher & 1,201 & 1,440 & 724 & 3,365 (5.0) \\
Male student & 609 & 720 & N/A & 1,329 (1.5) \\
Female student & 621 & 720 & N/A & 1,341 (1.7) \\
\hline
\hline
\end{tabular}
\end{table}

\begin{table}[t]
\centering
\caption{Number of utterances per each emotion}
\vspace{-3mm}
\label{tab:utterances_per_emotion}
\begin{tabular}{c|rrrr}
\hline
\hline
Speaker & Neutral & Happy & Sad & Angry \\
\hline
Teacher & 2,220 & 715 & 312 & 118 \\
Male student & 351 & 381 & 329 & 268 \\
Female student & 300 & 417 & 340 & 284 \\
\hline
\hline
\end{tabular}
\end{table}

\begin{table*}[t]
\centering
\caption{List of existing Japanese dialogue speech corpora}
\vspace{-3mm}
\label{tab:corpora}
\small
\begin{tabular}{cccccc}
\hline
\hline
Corpus & Dialogue type & Open-source & Dur~[hour] & \# of speakers & Emotion label \\
\hline
CSJ~\cite{maekawa00} & Interview/task-oriented/chitchat & No & 660 & 1,417 & No \\
Chiba3Party~\cite{den07} & Chitchat & No & 2 & 36 & No \\
UUDB~\cite{mori11} & Task-oriented & Yes & 2 & 14 & Yes \\
OGVC~\cite{arimoto12} & Voice chat while playing game & Yes & 14 & 17 & Yes \\
NICT-VADC~\cite{sugiura15} & Task-oriented & Yes & 7 & 1 & No \\
 \textbf{STUDIES (ours)} & Empathetic & Yes & 8 & 3 & Yes \\
\hline
\hline
\end{tabular}
\end{table*}

\vspace{-5pt}
\section{Corpus Analysis and Discussion}
\vspace{-3pt}
\subsection{Crowdsourcing setting and results}
\vspace{-3pt}
We used Lancers\footnote{https://www.lancers.jp/} as the crowdsourcing platform for collecting dialogue lines. The overall period of the crowdsourcing ran from June to July 2021. For the macrotask crowdsourcing to collect the long-dialogue lines, we employed 15 individual crowdworkers with writing experience. We asked each of the crowdworkers to write ten dialogue lines (four for neutral and two each for happy, sad, and angry initial emotions). We paid each of the crowdworkers approximately \$250 as compensation. We also started 12 microtasks (four each for happy, sad, and angry initial emotions) for creating the short-dialogue lines and recruited 100 crowdworkers (with overlap per task) for each microtask. We paid each of the crowdworkers approximately \$0.45 as compensation. As a result of our screening, we obtained 720 short-dialogue lines.

\vspace{-3pt}
\subsection{Corpus specification}\label{subsect:spec}
\vspace{-3pt}
Table~\ref{tab:utterances_per_subset} shows the number of utterances per subset of the STUDIES corpus. The total amount of data was about 5 hours for the teacher (about 4.3 hours excluding the ITA subset, phonetically balanced open-source sentences~\cite{ita}), and about 8.2 hours for the entire corpus.

Table~\ref{tab:utterances_per_emotion} shows the number of utterances per emotion label. The number of utterances for each emotion label of the students is generally balanced, as the results of situation-grounded chat dialogue collection. In addition, the majority of the teacher's utterances were of the neutral emotion and the number of angry utterances was extremely small. This trend is natural because empathy is different from sympathy, where a speaker's feelings are shared with another person (i.e., identification), as explained in Section~\ref{sect:intro}.

\vspace{-3pt}
\subsection{Comparison with existing corpora}
\vspace{-3pt}
Table~\ref{tab:corpora} lists existing Japanese dialogue speech corpora that can be used for the development of dialogue speech synthesis. Our STUDIES corpus advocates a new task called ``empathetic dialogue speech synthesis,'' which aims to synthesize speech that is empathetic with the interlocutor by considering his/her emotion in mind. In addition, the amount of data per speaker in our corpus exceeds one hour, which are larger than existing emotion-labeled dialogue speech corpora such as UUDB~\cite{mori11}, JTES~\cite{takeishi16jtes} and OGVC~\cite{arimoto12}.

\begin{figure}[t]
  \centering
  \includegraphics[width=0.9\linewidth, clip]{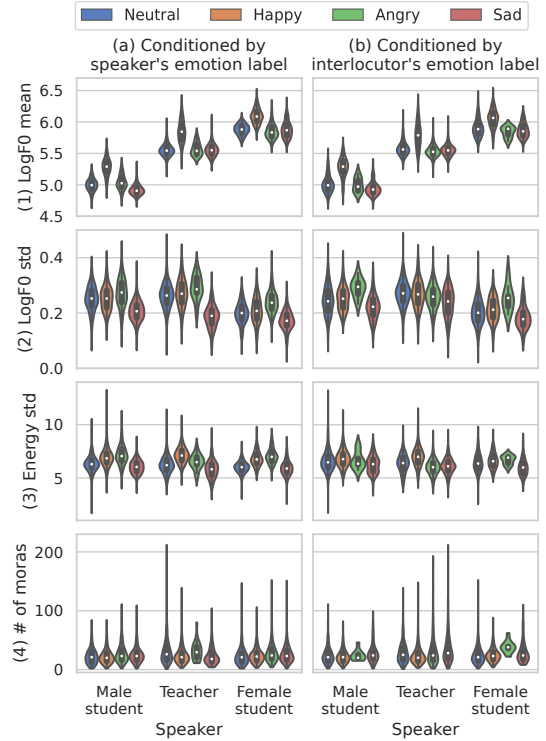}
  \vspace{-8pt}
  \caption{Prosodic feature statistics of STUDIES corpus.}
  \label{fig:prosody_stats}
\end{figure}

\vspace{-5pt}
\begin{table}[t]
\centering
\caption{Number of dialogues in training/validation/evaluation sets of STUDIES corpus we split. The number in parentheses means the number of utterances by the teacher}
\label{tab:teacher_data_split}
\vspace{-3mm}
\begin{tabular}{c|rrrr}
\hline
\hline
Set & & Long & & Short \\
\hline
Training & 126 & (1,743) & 600 & (1,200) \\
Validation & 12 & (91) & 60 & (120) \\
Evaluation & 12 & (91) & 60 & (120) \\
\hline
\hline
\end{tabular}
\end{table}

\vspace{-3pt}
\subsection{Analysis of prosodic features}
\vspace{-3pt}
To analyze the prosodic features of the empathetic dialogues, we calculated the mean and standard deviation (std) of the log $F_0$, std of energy, and the number of moras per utterance. We used WORLD~\cite{morise16world,morise16d4c} for the $F_0$ analysis.

Figure~\ref{fig:prosody_stats} shows the violin plot of the prosodic feature statistics. Here, columns (a) and (b) in this figure refer to the cases where the statistics were aggregated using the speaker's current emotion label and the interlocutor's last one, respectively. The former is aggregated on the basis of the emotion of the utterance itself, as in normal speech emotion, while the latter is considered to be the results aggregated as a response to the interlocutor. Rows (1)--(4) in this figure indicate each statistics. From the results shown in Fig.~\ref{fig:prosody_stats}, we can observe the following three trends in the prosodic features of the teacher's empathetic dialogue speech. First, the mean of log $F_0$ (Fig.~\ref{fig:prosody_stats}(1b)) does not differ in general trend from the result aggregated by the teacher's own emotion label (Fig.~\ref{fig:prosody_stats}(1a)). This result suggests that the rough trends in prosodic features of empathetic dialogue speech can be predicted if the interlocutor's emotion can be estimated. Second, the standard deviations of log $F_0$ and energy (Figs.~\ref{fig:prosody_stats}(2b)--(3b)) get smaller and larger than those as shown in Figs.~\ref{fig:prosody_stats}(2a)--(3a) when the student's emotions are angry and sad, respectively. This is the opposite of the results obtained when the teacher's emotion labels were used for aggregating the calculated results (Figs.~\ref{fig:prosody_stats}(2a)--(3a)), which indicates that empathy is actually a different concept from sympathy. Finally, the number of moras per utterance (Fig.~\ref{fig:prosody_stats}(4b)) tends to become large when the student is feeling negative (i.e., angry or sad). These results suggest that the use of the interlocutor's emotion label alone is insufficient to reproduce the variability of prosodic features in the empathetic dialogue speech synthesis, and that rich additional information, such as the context of the dialogue, may be necessary.

\vspace{-5pt}
\section{TTS Experiment}
\vspace{-3pt}
\subsection{Experimental conditions}
\vspace{-3pt}
We built the voice agent of the teacher in our STUDIES corpus and evaluated how natural the agent's voice was. We split the teacher's speech data into the long/short dialogue subsets for training, validation, and evaluation sets as shown in Table~\ref{tab:teacher_data_split}. We divided the validation and evaluation so that the gender of the students and three initial emotions (happy, angry, and sad) were balanced. We used all 724 utterances in the ITA subset as training data. We downsampled the speech data to 22,050~Hz.

We used FastSpeech 2~\cite{ren21} as the acoustic model for neural TTS, with the PyTorch implementation for Japanese TTS\footnote{https://github.com/Wataru-Nakata/FastSpeech2-JSUT}. We followed the settings of a deep neural network (DNN) architecture and speech parameter extraction of this implementation. We used the WORLD vocoder~\cite{morise16world,morise16d4c} to estimate $F_0$. We modified the variance adaptors to predict the speech features averaged over phonemes, referring to FastPitch~\cite{lancucki21} for stable training and higher synthetic speech quality. The optimization algorithm was Adam~\cite{kingma14adam} with an initial learning rate $\eta$ of 0.0625, $\beta_1$ of 0.9, and $\beta_2$ of 0.98.

We used the HiFi-GAN vocoder~\cite{kong20} for speech waveform generation. We first pretrained HiFi-GAN using the JSUT corpus~\cite{takamichi20ast} with 100 epochs. We then fine-tuned it by using the same training data as that for FastSpeech 2 (shown in Table~\ref{tab:teacher_data_split}) with 200 epochs. The optimization algorithm was Adam with an initial learning rate $\eta$ of 0.0003, $\beta_1$ of 0.8, and $\beta_2$ of 0.99.

\begin{figure}[t]
  \centering
  \includegraphics[width=0.99\linewidth, clip]{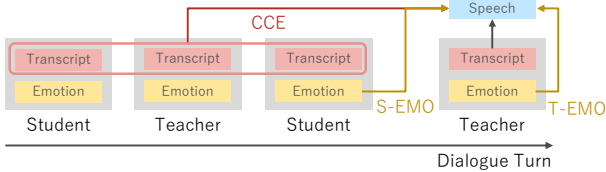}
  \vspace{-8pt}
  \caption{Information used in the TTS experiment. The ground-truth dialogue history (text and emotion labels) is fed into the CCE to extract conversational context embedding. The emotion labels are converted to emotion embeddings through a trainable lookup embedding table.}
  \label{fig:6methods}
\end{figure}

We compared the following six models in this experiment:
\vspace{-2.5pt}
\begin{itemize}
    \leftskip -5mm \itemsep -0mm
\item {\bf FS2}: Original FastSpeech 2~\cite{ren21}
\item {\bf FS2\_T-EMO}: FS2 conditioned by emotion label of the teacher's current utterance
\item {\bf FS2\_S-EMO}: FS2 conditioned by emotion label of male/female student's last utterance
\item {\bf FS2\_CCE}: FS2 with conversational context encoder (CCE)~\cite{guo21}
\item {\bf FS2\_CCE\_T-EMO}: FS2\_CCE conditioned by emotion label of the teacher's current utterance
\item {\bf FS2\_CCE\_S-EMO}: FS2\_CCE conditioned by emotion label of male/female student's last utterance
\end{itemize}
\vspace{-2.5pt}
Figure~\ref{fig:6methods} illustrates these methods. The CCE extracts a conversational context embedding from joint vectors of emotion embedding sequences through a trainable lookup table and sentence embedding sequences obtained by using BERT~\cite{devlin19} pretrained by Japanese text data\footnote{https://huggingface.co/colorfulscoop/sbert-base-ja}. We used a publicly available PyTorch implementation of the CCE\footnote{https://github.com/keonlee9420/Expressive-FastSpeech2/tree/conversational} and modified it to extract 256-dimensional context embeddings from the sentence embeddings of up to ten previous utterances in one dialogue. In addition to the six models, we also compared ``Reconst.'' (i.e., analysis-synthesized speech by using the HiFi-GAN vocoder).

\vspace{-3pt}
\subsection{Subjective evaluations}
\vspace{-3pt}
We conducted a five-scaled mean opinion score (MOS) test on utterance-level naturalness of synthetic speech. We presented 35 synthetic speech samples to listeners in random order. Listeners rated the naturalness of each speech sample from degrees of 1 (``very unnatural'') to 5 (``very natural''). We recruited 100 listeners using our crowdsourcing subjective evaluation system, and we collected 3,500 answers.

We also conducted a five-scaled MOS test on {\it dialogue-level} naturalness of synthetic speech to investigate what kind of information is critical for reproducing the teacher's empathetic speech. We presented listeners with 35 dialogue speech samples taken from the short-dialogue lines, which consisted of two natural speech samples of a male/female student and two synthetic speech samples of the teacher generated by each of the seven methods (six models and ``Reconst.'') in random order. We also presented listeners with the dialogue purpose (see Section \ref{subsect:scenario}) and asked them to evaluate the teacher's voice regarding the naturalness as a response to the student. We did not limit the number of times the listeners could listen to the sample, but we instructed them to listen intently to all four voices in a presented dialogue. We recruited 100 listeners using our crowdsourcing subjective evaluation system, and we collected 3,500 answers.

Table~\ref{tab:mos} shows the evaluation results. From the results of the utterance-level naturalness evaluation, there was no significant difference between the MOSs of the six TTS models, and only ``Reconst.'' achieved a MOS higher than 4. From the results of the dialogue-level naturalness evaluation, three models, ``FS2\_T-EMO,'' ``FS2\_CCE,'' and ``FS2\_CCE\_S-EMO,'' achieved a MOS significantly higher than ``FS2.'' These results suggest that high-quality empathetic dialogue speech synthesis can be achieved when the emotion label of the speaker or the text information of the dialogue history is available. Furthermore, even if the speaker's emotion label is unavailable, natural empathetic dialogue speech synthesis can be achieved by combining the interlocutor's emotion label with the dialogue history information. The speech samples used in the subjective evaluations are available at http://sython.org/Corpus/STUDIES/demo\_STUDIES.html.

\begin{table}[t]
\centering
\caption{Results of MOS tests on speech naturalness with their 95\% confidence intervals. \textbf{Bold} scores are significantly higher than those of ``FS2'' ($p<0.05$)}
\label{tab:mos}
\vspace{-3mm}
\begin{tabular}{lrr}
\hline
\hline
Method & Utterance-level & Dialogue-level  \\
\hline
Reconst. & 4.08$\pm$0.09 & 4.33$\pm$0.08 \\
FS2 & 3.42$\pm$0.10 & 3.20$\pm$0.11 \\
FS2\_T-EMO & 3.43$\pm$0.10 & \textbf{3.38$\pm$0.11} \\
FS2\_S-EMO & 3.38$\pm$0.10 & 3.29$\pm$0.11 \\
FS2\_CCE & 3.32$\pm$0.10 & \textbf{3.37$\pm$0.10} \\
FS2\_CCE\_T-EMO & 3.38$\pm$0.10 & 3.34$\pm$0.10 \\
FS2\_CCE\_S-EMO & 3.43$\pm$0.10 & \textbf{3.41$\pm$0.11} \\
\hline
\hline
\end{tabular}
\end{table}

\vspace{-5pt}
\section{Conclusion}
\vspace{-3pt}
We introduced the STUDIES corpus, which we newly constructed to advance friendly TTS technologies. We described our methodology to construct an empathetic dialogue speech corpus and conducted a TTS experiment for reproducing an empathetic speaking style. The results demonstrated that the interlocutor's emotion label and conversational context embedding produced speech with the same degree of naturalness as that synthesized by using the speaker's emotion label. In the future, we will collect additional empathetic dialogue data and analyze potential bias of our corpus. We also plan to conduct human study using our empathetic dialogue TTS system.

\textbf{Acknowledgements:}
This research was conducted as joint research between LINE Corporation and Saruwatari-Koyama Laboratory of The University of Tokyo, Japan.

   \ninept
   \bibliographystyle{IEEEtran}
   \bibliography{template}

\begin{thebibliography}{10}
\providecommand{\url}[1]{#1}
\csname url@samestyle\endcsname
\providecommand{\newblock}{\relax}
\providecommand{\bibinfo}[2]{#2}
\providecommand{\BIBentrySTDinterwordspacing}{\spaceskip=0pt\relax}
\providecommand{\BIBentryALTinterwordstretchfactor}{4}
\providecommand{\BIBentryALTinterwordspacing}{\spaceskip=\fontdimen2\font plus
\BIBentryALTinterwordstretchfactor\fontdimen3\font minus
  \fontdimen4\font\relax}
\providecommand{\BIBforeignlanguage}[2]{{%
\expandafter\ifx\csname l@#1\endcsname\relax
\typeout{** WARNING: IEEEtran.bst: No hyphenation pattern has been}%
\typeout{** loaded for the language `#1'. Using the pattern for}%
\typeout{** the default language instead.}%
\else
\language=\csname l@#1\endcsname
\fi
#2}}
\providecommand{\BIBdecl}{\relax}
\BIBdecl

\bibitem{sagisaka88}
Y.~Sagisaka, ``Speech synthesis by rule using an optimal selection of
  non-uniform synthesis units,'' in \emph{Proc. ICASSP}, New York, U.S.A., Apr.
  1988, pp. 679--682.

\bibitem{shen18}
J.~Shen, R.~Pang, R.~J. Weiss, M.~Schuster, N.~Jaitly, Z.~Yang, Z.~Chen,
  Y.~Zhang, Y.~Wang, R.~Skerry-Ryan, R.~A. Saurous, Y.~Agiomyrgiannakis, and
  Y.~Wu, ``Natural {TTS} synthesis by conditioning {W}ave{N}et on mel
  spectrogram predictions,'' in \emph{Proc. ICASSP}, Calgary, Canada, Apr.
  2018, pp. 4779--4783.

\bibitem{kim21vits}
J.~Kim, J.~Kong, and J.~Son, ``Conditional variational autoencoder with
  adversarial learning for end-to-end text-to-speech,'' in \emph{Proc. ICML},
  Virtual Conference, Jun. 2021, pp. 5530--5540.

\bibitem{lee17nips}
Y.~Lee, A.~Rabiee, and S.-Y. Lee, ``Emotional end-to-end neural speech
  synthesizer,'' in \emph{Proc. NIPS}, Long Beach, U.S.A., Dec. 2017.

\bibitem{wang18gst}
Y.~Wang, D.~Stanton, Y.~Zhang, R.~Skerry-Ryan, E.~Battenberg, J.~Shor, Y.~Xiao,
  F.~Ren, Y.~Jia, and R.~A. Saurous, ``Style {T}okens: Unsupervised style
  modeling, control and transfer in end-to-end speech synthesis,'' in
  \emph{Proc. ICML}, Stockholm, Sweden, Jul. 2018, pp. 5167--5176.

\bibitem{li21}
X.~Li, C.~Song, J.~Li, Z.~Wu, J.~Jia, and H.~Meng, ``Towards multi-scale style
  control for expressive speech synthesis,'' in \emph{Proc. INTERSPEECH}, Sep.
  2021, pp. 4673--4677.

\bibitem{hojat09}
M.~Hojat, ``Ten approaches for enhancing empathy in health and human services
  cultures,'' \emph{Journal of Health and Human Services Administration},
  vol.~31, no.~4, pp. 412--450, 2009.

\bibitem{warren15}
C.~Warren and B.~K. Hotchkins, ``Teacher education and the enduring
  significance of ``false empathy'','' \emph{The Urban Review}, vol.~47, no.~2,
  pp. 266--292, Jun. 2015.

\bibitem{dabis96}
M.~H. Davis, \emph{Empathy: {A} Social Psychological Approach}.\hskip 1em plus
  0.5em minus 0.4em\relax Routledge, 1996.

\bibitem{rashkin19}
H.~Rashkin, E.~M. Smith, M.~Li, and Y.-L. Boureau, ``Towards empathetic
  open-domain conversation models: a new benchmark and dataset,'' in
  \emph{Proc. ACL}, Florence, Italy, Aug. 2019, pp. 5370--5381.

\bibitem{li20emodg}
Q.~Li, H.~Chen, Z.~Ren, P.~Ren, Z.~Tu, and Z.~Chen, ``{EmpDG}: Multi-resolution
  interactive empathetic dialogue generation,'' in \emph{Proc. COLING},
  Barcelona, Spain, Dec. 2020, pp. 4454--4466.

\bibitem{regenbogen12}
C.~Regenbogen, D.~A. Schneider, A.~Finkelmeyer, N.~Kohn, B.~Derntl,
  T.~Kellermann, R.~E. Gur, F.~Schneider, and U.~Habel, ``The differential
  contribution of facial expressions, prosody, and speech content to empathy,''
  \emph{Cognition and Emotion}, vol.~26, no.~6, pp. 995--1014, Jan. 2012.

\bibitem{mctear20}
M.~McTear, \emph{Conversational {AI}: Dialogue Systems, Conversational Agents,
  and Chatbots}.\hskip 1em plus 0.5em minus 0.4em\relax Morgan \& Claypool,
  2020.

\bibitem{zhang18acl}
S.~Zhang, E.~Dinan, J.~Urbanek, A.~Szlam, D.~Kiela, and J.~Weston,
  ``Personalizing dialogue agents: {I} have a dog, do you have pets too?'' in
  \emph{Proc. ACL}, Melbourne, Australia, Jul. 2018, pp. 2204--2213.

\bibitem{li16acl}
J.~Li, M.~Galley, C.~Brockett, G.~Spithourakis, J.~Gao, and B.~Dolan, ``A
  persona-based neural conversation model,'' in \emph{Proc. ACL}, Berlin,
  Germany, Aug. 2016, pp. 994--003.

\bibitem{vinyals15}
O.~Vinyals and Q.~Le, ``A neural conversational model,'' in \emph{Proc. ICML
  {D}eep {L}earning {W}orkshop}, Lille, France, Jul. 2015.

\bibitem{simperl15}
E.~Simperl, ``How to use crowdsourcing effectively: Guidelines and examples,''
  \emph{LIBER Quarterly}, vol.~25, no.~1, pp. 18--39, Aug. 2015.

\bibitem{dawid79}
A.~P. Dawid and A.~M. Skene, ``Maximum likelihood estimation of observer
  error-rates using the {EM} algorithm,'' \emph{Applied Statistics}, vol.~28,
  no.~1, pp. 20--28, May 1979.

\bibitem{ita}
J.~Koguchi, I.~Kanai, Y.~Oda, T.~Saito, and M.~Morise, ``{ITA} {C}orpus,''
  \url{https://github.com/mmorise/ita-corpus}, 2021.

\bibitem{maekawa00}
K.~Maekawa, H.~Koiso, S.~Furui, and H.~Isohara, ``Spontaneous speech corpus of
  {J}apanese,'' in \emph{Proc. LREC}, Athens, Greece, May 2000, pp. 947--952.

\bibitem{den07}
Y.~Den and M.~Enomoto, \emph{Conversational Informatics: An Engineering
  Approach}.\hskip 1em plus 0.5em minus 0.4em\relax John Wiley \& Sons, 2007.

\bibitem{mori11}
H.~Mori, T.~Satake, M.~Nakamura, and H.~Kasuya, ``Constructing a spoken
  dialogue corpus for studying paralinguistic information in expressive
  conversation and analyzing its statistical/acoustic characteristics,''
  \emph{Speech Communication}, vol.~53, pp. 57--65, Jan. 2011.

\bibitem{arimoto12}
Y.~Arimoto, H.~Kawatsu, S.~Ohno, and H.~Iida, ``Naturalistic emotional speech
  collection paradigm with online game and its psychological and acoustical
  assessment,'' \emph{Acoustical Science and Technology}, vol.~33, no.~6, pp.
  359--369, Jun. 2012.

\bibitem{sugiura15}
K.~Sugiura, Y.~Shiga, H.~Kawai, T.~Misu, and C.~Hori, ``A cloud robotics
  approach towards dialogue-oriented robot speech,'' \emph{Advanced Robotics},
  vol.~29, no.~7, pp. 449--456, Mar. 2015.

\bibitem{takeishi16jtes}
E.~Takeishi, T.~Nose, Y.~Chiba, and A.~Ito, ``Construction and analysis of
  phonetically and prosodically balanced emotional speech database,'' in
  \emph{Proc. Oriental COCOSDA}, Bali, Indonesia, Oct. 2016, pp. 16--21.

\bibitem{morise16world}
M.~Morise, F.~Yokomori, and K.~Ozawa, ``{WORLD}: a vocoder-based high-quality
  speech synthesis system for real-time applications,'' \emph{IEICE
  Transactions on Information and Systems}, vol. E99-D, no.~7, pp. 1877--1884,
  2016.

\bibitem{morise16d4c}
M.~Morise, ``{D4C}, a band-aperiodicity estimator for high-quality speech
  synthesis,'' \emph{Speech Communication}, vol.~84, pp. 57--65, Nov. 2016.

\bibitem{ren21}
Y.~Ren, C.~Hu, X.~Tan, T.~Qin, S.~Zhao, Z.~Zhao, and T.-Y. Liu, ``{FastSpeech}
  2: Fast and high-quality end-to-end text to speech,'' in \emph{Proc. ICLR},
  Vienna, Austria, May 2021.

\bibitem{lancucki21}
A.~Łańcucki, ``Fast{P}itch: Parallel text-to-speech with pitch prediction,''
  in \emph{Proc. ICASSP}, Montreal, Canada, Jun. 2021, pp. 6588--6592.

\bibitem{kingma14adam}
D.~Kingma and B.~Jimmy, ``Adam: A method for stochastic optimization,'' in
  \emph{ar{X}iv preprint ar{X}iv:1412.6980}, 2014.

\bibitem{kong20}
J.~Kong, J.~Kim, and J.~Bae, ``{HiFi-GAN}: Generative adversarial networks for
  efficient and high fidelity speech synthesis,'' in \emph{Proc. NeurIPS},
  Vancouver, Canada, Dec. 2020.

\bibitem{takamichi20ast}
S.~Takamichi, R.~Sonobe, K.~Mitsui, Y.~Saito, T.~Koriyama, N.~Tanji, and
  H.~Saruwatari, ``{JSUT and JVS}: Free {J}apanese voice corpora for
  accelerating speech synthesis research,'' \emph{Acoustical Science and
  Technology}, vol.~41, no.~5, pp. 761--768, Sep. 2020.

\bibitem{guo21}
H.~Guo, S.~Zhang, F.~K. Soong, L.~He, and L.~Xie, ``Conversational end-to-end
  {TTS} for voice agent,'' in \emph{Proc. SLT}, Shenzhen, China, Jan. 2021, pp.
  403--409.

\bibitem{devlin19}
J.~Devlin, M.-W. Chang, K.~Lee, and K.~Toutanova, ``{BERT}: Pre-training of
  deep bidirectional transformers for language understanding,'' in \emph{Proc.
  NAACL-HLT}, Minneapolis, U.S.A., Jun. 2019, pp. 4171--4186.

\end{thebibliography}

\end{document}